# EPA-RIMM: A Framework for Dynamic SMM-based Runtime Integrity Measurement

Brian Delgado, Karen Karavanic
Portland State University
bdelgado[†], karavan@cs.pdx.edu

## ABSTRACT

Runtime integrity measurements identify unexpected changes in operating systems and hypervisors during operation, enabling early detection of persistent threats. System Management Mode, a privileged x86 CPU mode, has the potential to effectively perform such rootkit detection. Previously proposed SMM-based approaches demonstrated effective detection capabilities, but at a cost of performance degradation and software side effects. In this paper we introduce our solution to these problems, an SMM-based Extensible, Performance Aware Runtime Integrity Measurement Mechanism called EPA-RIMM. The EPA-RIMM architecture features a performance-sensitive design that decomposes large integrity measurements and schedules them to control perturbation and side effects. EPA-RIMM's decomposition of long-running measurements into shorter tasks, extensibility, and use of SMM complicates the efforts of malicious code to detect or avoid the integrity measurements. Using a Minnowboard-based prototype, we demonstrate its detection capabilities and performance impacts. Early results are promising, and suggest that EPA-RIMM will meet production-level performance constraints while continuously monitoring key OS and hypervisor data structures for signs of attack.

## KEYWORDS

BIOS; SMM; firmware; rootkit detection; runtime integrity measurement; security; performance

## 1 INTRODUCTION

Today's complex server platforms include software environments vulnerable to sophisticated malware called *rootkits*. Rootkits target sensitive low-level kernel or hypervisor data structures - such as interrupt handlers, event handlers, registers, and memory - remaining undetected for extended periods to achieve persistence. One response to this vulnerability has been the development of runtime integrity measurement mechanisms (RIMMs) that aim to detect rootkits before financial, political, or other damage occurs. The runtime integrity measurement approach periodically preempts execution and examines the interrupted state, checking for unexpected changes in low-level resources. Several approaches leveraging a privileged CPU mode called System Management Mode (SMM) to perform the needed monitoring of resources in the runtime environment demonstrated promising rootkit detection capabilities [3][4][5].

SMM is a general purpose, widely available mechanism on x86 CPUs. Firmware on computer systems performs a variety of critical management tasks at runtime using SMM. This can include managing CPU power states, controlling low-level hardware, handling thermal throttling, performing BIOS flash updates [34], and handling memory errors, among other usages[1]. SMM operates at a more privileged layer than host-side kernel or hypervisor code. The entry into SMM is accomplished by a System Management Interrupt (SMI) which typically takes all CPU threads[1] out of the operating system or applications and into the SMM handler. However, the SMI's work is generally processed by a single CPU thread. SMIs can be triggered by Ring 0 software by writing values to the APM_CNT IO port on Intel platforms [46] and the "SMI Command Port" on AMD platforms [47]. Intel's Itanium processors also feature a Platform Management Interrupt (PMI) that is similar to an SMI [49]. A proposal has been made to bring SMM-like functionality to ARM platforms utilizing TrustZone [48]. Upon entering SMM, the SMRAM Save State Map is populated with CPU register values from the time of interruption. An SMM RIMM can inspect this data to determine if unexpected changes have occurred. SMM code issues an RSM instruction to exit SMM and resume host-side execution at the point of interruption. The Intel BITS tool measures the duration of SMIs on a system and warns if they are larger than "acceptable" limits (e.g. 150 microseconds) [7].

For RIMM purposes, SMM provides critical benefits:

- SMM runs at a higher privilege than host software;
- SMM has broad visibility into operating system and hypervisor (host software) memory[2] and CPU registers
- SMM is strongly isolated from host software as the hardware-protected SMRAM can only be read or written from SMM, providing a location to place measurement software
- When an SMI occurs, the transitions into SMM cannot be prevented by malicious code;
- When all CPU threads enter SMM, host software execution is paused which presents the opportunity to inspect the system while it is temporarily halted; and
- As x86 systems broadly support SMM, the mechanisms can be readily adopted.

---

[1] As some Intel CPUs feature the HyperThreading feature which creates a set of logical CPUs, we refer to the logical CPUs as "CPU threads."

[2] Recent open-source UEFI code updates have constrained SMM's memory visibility. We discuss the implications and potential options in Section 5.

---

† This author was a full-time employee of Intel Corp. when this work was done.

For these reasons, SMM RIMMs present intriguing possibilities for adding new detection capabilities to combat host software rootkits.

Our research demonstrated that an important drawback limits the potential of SMM-RIMMs: interference with system software that may lead to significant perturbation of the system and even software failures [2]. System software assumptions regarding scheduling regularity as well as task durations are challenged by prolonged periods of time in SMM. Early SMM RIMMs exceeded the published SMM guidelines by orders of magnitude. There are other limitations to early SMM-based approaches: the lack of extensibility - the inability to dynamically change the set of monitored resources to maintain effectiveness as new rootkits emerge; and the inability to adjust the frequency of checking as new threat information becomes available.

In this paper, we present our solution to this problem, an SMM-RIMM framework called EPA-RIMM. EPA-RIMM's goal is to provide quick detection of kernel or hypervisor rootkits by identifying unexpected changes in system state snapshots. It accomplishes this by periodically interrupting the running system to inspect sets of presumed static resources to identify changes, any of which would be a strong indicator of compromise.

The key contributions of this work are:

1. *A mechanism for decomposing large integrity measurements to remain consistent with expectations regarding SMI latency*. This approach removes the significant performance degradation incurred by prolonged SMIs and creates an opportunity for the development of new integrity measurements that were formerly impractical.

2. *A mechanism for throttling the rate of integrity checking*. EPA-RIMM's scheduler facilitates a varying time budget for measurements, allowing the performance-security tradeoff to be adjusted during runtime based on the threat landscape. This allows EPA-RIMM to reduce its performance impact on the system when necessary but also supports increasing it as threats emerge.

3. *An API to specify measurements during operation*. This allows the measurements to vary depending on the environment and abstracts the complexity of OS/VMM-specific details from the lowest level SMM-code.

4. *An open-source prototype of EPA-RIMM to demonstrate its effectiveness and performance*. We plan to release EPA-RIMM as open-source software to facilitate its use as a research and educational tool.

EPA-RIMM supports either native Linux or virtualized systems. In this paper, we introduce EPA-RIMM and focus on its use for native Linux.

## 2   EPA-RIMM Architecture

This section details the EPA-RIMM Architecture as shown in Figure 1. EPA-RIMM comprises the Diagnosis Manager (Section 2.1), the Backend Manager (Section 2.2), the Host Communications Manager (Section 2.3) and the Inspector

(Section 2.4). We discuss the architecture in Section 2.5. A detailed example is provided in Appendix A.

There are three key abstractions: Checks, Tasks, and Bins. A *Check* is a description of an integrity measurement, including a command and its arguments, a priority, and a decomposition target (to guide decomposition). Checks allow the Administrator to specify particular measurements over sets of memory regions, Control Registers, and Model-Specific Registers (MSRs). Sample Checks include: "Static Linux Kernel Code Sections" that measures the Linux kernel code sections to identify code injections, the "IDTR" Check that verifies that the IDTR register value has not unexpectedly changed, the "GDT" Check that measures the Global Descriptor Table (GDT) to determine if it has changed. Other Checks measure specific MSRs or CPU control registers (e.g. CR0, CR4), for example, to determine if the Supervisor Mode Execution Protection (SMEP) were disabled by malware. At runtime, Checks are decomposed into some number of *Tasks*, or partial resource measurements to meet expectations for SMI latency. Tasks are scheduled by filling *Bins* which consist of the set of work to be performed in one SMI session. Each Bin's size is defined as the sum of the execution times of the Tasks it contains.

### 2.1   Diagnosis Manager

The Diagnosis Manager (DM) is the component that orchestrates the runtime integrity measurements. It decides which inspections to run under at a given point of time and interprets the measurement results. A single DM may be responsible for one or more nodes in a cluster, and may communicate with other Diagnosis Managers. The DM sends and receives information about attack discoveries from across the EPA-RIMM framework to help guide detection on other monitored nodes. This allows dynamically adjusting the priority of measurements to search for detected issues on other nodes. The DM can also interface with an external Security Information and Event Management system (SIEM) to send and receive telemetry on attacks, although a full description of this interface is outside of the scope of this paper.

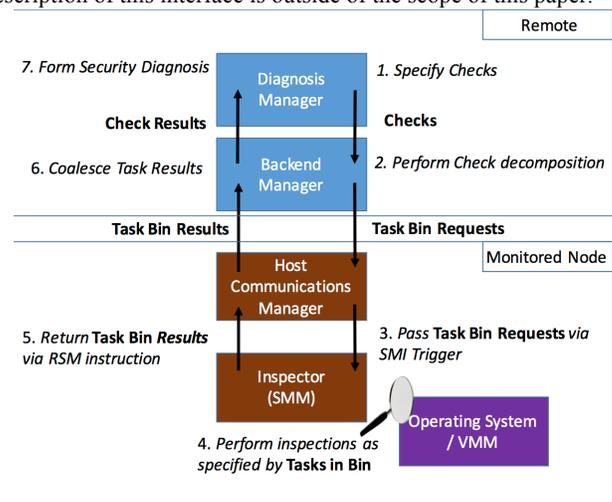

**Figure 1 EPA-RIMM Architecture**

*Provisioning*. An EPA-RIMM Administrator initializes the Diagnosis Manager with a set of specific Checks. Each Check Description (left column of **Table 1**) describes the measurement operation in sufficient detail for the Backend Manger to decompose the Check into Tasks and for the Inspector to process the measurement work, including the command and parameters. **Table 2** describes the currently supported EPA-RIMM commands along with their accompanying parameters. Checks that measure large memory regions could exceed desired SMI session times and need decomposition. To determine a suitable granularity, a simple micro-benchmark is run during the provisioning phase to measure the cost per byte of various hash sizes. These parameters combined with the Check's Decomposition Target in microseconds determines the appropriate granularity. Checks involving Control Registers or MSRs cannot be decomposed and thus consist of a single Task. **Table 4** shows the parameters that govern the Check decomposition.

*Runtime*. The DM sends Checks to the Backend Manager. Each Check returns true or false, indicating whether the measurement matches the comparison value. "False" results cause the DM to raise an alert. The Diagnosis Manager has the option of increasing or decreasing the threat level, by notifying both the Backend Manager and any connected Diagnosis Managers.

**Table 1 Check and Task Descriptions**

| Check | Task | Description |
|---|---|---|
| ID # | | Unique ID |
| Command | | Measurement to perform |
| Operand | | Command arg |
| Memory Address | | Starting address |
| Length | | Memory Measurement Size |
| Last inspection time | | |
| Priority | | |
| | Nonce Value | Liveness indicator |
| Diagnosis Manager Signature | Backend Manager Signature | |
| | Comparison Hash | Value to compare |
| Decomposition Target | | |

**Table 2 Measurement Commands and Selected Parameters**

| Command | Selected Parameters |
|---|---|
| Register | CR0,CR3*,CR4,IDTR,GDTR |
| (Virt\|phys) Mem | Address, Length |
| MSR | MSR Number |

## 2.2 Backend Manager

The Backend Manager (BEM) manages the performance aspects of EPA-RIMM and provides measurement requests to the monitored systems. It receives Checks from the Diagnosis Manager and decomposes them into smaller Tasks to avoid prolonged SMM session times. The Backend Manager schedules tasks by filling Bins based on a *target Bin size*. It signs, creates a MAC, and encrypts each Bin and then provides it to the Host Communications Manager which interfaces with the SMM-based Inspector. The BEM waits to receive the Inspector's results back. It decrypts the results and checks the signature and MAC to ensure that they came from the proper Inspector and were not tampered with in-transit. The BEM merges individual Task results into a single *Check* result of true or false and sends the results to the Diagnosis Manager.

*Provisioning*. The Backend Manager is provisioned with the appropriate keys for signing and encryption. The Backend Manager's Task performance estimations are set based on an initial performance measurement so that the EPA-RIMM can begin runtime operation with an appropriately sized amount of work in a Bin. Since there is no preemption, platform-specific performance prediction is accurate for SMM code. Initial measurements that will be re-checked over time can be gathered in the provisioning phase in an offline environment (preferred) or upon the first measurement of a resource. The latter measurement option is less desirable due to the possibility that malware could have an opportunity to register initial values.

*Runtime*. As the system experiences an overhead transitioning to SMM and back, minimizing the number of SMIs becomes a consideration. Since each SMI transfers a Bin, efficiently filling the Bin reduces the number of SMIs and consequently the amount of time spent transitioning to and from SMM. We therefore use a backpack algorithm:

*Given a set of n Tasks, select a subset to fill a bounded-size Bin with max cost C, where each Task i has an assigned cost, $c_i$, and an assigned value, $v_i$.*

A set of Tasks can only be scheduled into the same Bin if the sum of their costs is less than or equal to the max cost C. The values represent priorities, and the costs represent the estimated runtime on a given system. Thus, tasks are not scheduled in strict priority order; a lower priority Task might be selected to "fill" a Bin in place of larger higher priority tasks. Priorities are adaptive; tasks are assigned an initial priority based on their parent Check, but priorities change at runtime for example with aging. The BEM may increase or decrease the Bin size within set bounds. For example, it may set the Bin size to the *maximum Bin size Cmax* during times of elevated threat. *Cmax* provides an upper bound to the amount of time spent in SMM in a single session to maintain adequate system performance. The BEM may also increase or decrease the SMI frequency upon direction from the DM.

## 2.3 Host Communication Manager (HCM)

The Host Communication Manager resides within the monitored system and provides an interface between the Inspector and the BEM. The HCM should be implemented via an out-of-band mechanism such as the BMC (Baseboard Management

Controller) [3]. In-band mechanisms (e.g. Ring 3 application and kernel module) should not be used as they are vulnerable to malware. For example, if a Host Communication Manager process were to be killed by malware, the measurement would stop. While the Backend Manager could detect a lack of response and trigger an alert, there is a subtler attack that is possible. Malicious code could recognize that a measurement request is imminent, clean its traces, and then let the measurement proceed. The HCM receives bins from the BEM and provides these bins to the Inspector when it triggers its operation via an SMI. The HCM also receives results from the Inspector which it returns to the BEM.

## 2.4 The Inspector

The Inspector performs the actual runtime integrity measurements from SMM, noting differences in the current measurement compared to the comparison measurement. The Inspector is compiled into the BIOS and is initiated via an SMI. It has the ability to view the interrupted host-side CPU register state by checking the SMRAM Save State Map, examining MSR values, and measuring specified regions of the host-side memory space. The Inspector also monitors the amount of time required by the measurement and returns the cost to the Backend Manager so that it can adaptively tune the Bin size.

*Provisioning*. The Inspector must be provisioned with encryption and signing keys.

*Runtime*. At runtime, the Inspector will be invoked by an SMI coming from the Host Communication Manager that specifies the Bin that the Inspector should operate upon.

The virtual addresses of the Linux kernel in the System.map or /proc/kallsyms that list code or data locations cannot be directly used by the Inspector as it operates based on physical addresses. The SMM Inspector walks the page tables using the CR3 to convert virtual addresses to physical. This mechanism provides the ability to bridge between the virtual and physical addressing schemes [4]. The Inspector returns the results as shown in **Table 3.**

The kernel address space layout randomization (KASLR) feature would require special handling with EPA-RIMM measurements as the kernel addresses would be randomized. One option for supporting KASLR would be to generate new provisioned values upon initial boot. However, at present, the future of KASLR is unclear [52]. There have been several recent attacks on the KASLR feature using page faults, prefetch, Intel TSX, and Branch Target Buffers and several mitigations proposed [39][40][41][42].

## 2.5 Discussion

The design of EPA-RIMM has several features that reduce the SMM storage requirements. As it is infeasible to update the BIOS at runtime, the inspector supports a basic set of measurement commands as opposed to building in operating system or hypervisor-specific details that could change with host software updates. Storing the measurement hashes in the Backend Manager avoids scalability issues due to limited SMM storage space.

EPA-RIMM's Check decomposition feature has benefits beyond reducing system impacts; It allows measurements to become more frequent and less predictable. Frequent measurements reduce the time interval that malware has to operate without potentially being inspected. Additionally, decomposition allows measuring a wider variety of resources than would be feasible before.

EPA-RIMM gathers partial results over time instead of complete measurement results gathered atomically. However, scheduling security checks in a single atomic measurement session has one clear benefit; At a precise point in time, the measurement shows whether an entire resource is unchanged. The challenge with this approach is that security checks could take unbounded amounts of time and the number of potential checks grows over time. This results in clear scalability problems. Therefore, EPA-RIMM inspects more granular portions of resources over time, effectively presenting a rolling confidence metric.

Supporting an extensible Check specification interface helps prevent measurements from becoming outdated in the presence of evolving rootkit techniques. SMM RIMMs that build in extensive host-side context into an SMM-based measurement agent are not suited to dealing with quickly emerging threats and host-side software changes.

**Table 3 Results Description**

| Results Entry | Source |
|---|---|
| Check ID # | BEM |
| Task ID # | BEM |
| Result | Inspector |
| Measured Hash | Inspector |
| Measurement Cost | Inspector |
| Inspector Signature | Inspector |
| Results integrity measurement | Inspector |

**Table 4 Decomposition and Bin Size Parameters**

| Value | Configurable at… | Applies to… |
|---|---|---|
| Target Bin Size | Runtime (by Diagnosis Manager) | Current Bin size |
| Maximum Bin Size | Provisioning | Bin size limit |
| Decomposition Target | Provisioning | Task granularity |

However, there are challenges in adding new resources to monitor at runtime. If the set of measured items is not provisioned at host software initialization, it is possible that malware has already changed the value of one of the resources. Strategies to address this complication could consist of measuring resources on an offline system or comparing results from a set of identically configured monitored nodes. Malware that installed itself and later tried to remove itself could be detected via EPA-RIMM re-measurements.

All SMM RIMMs must address the scenario that once a resource has been checked, the measurement could go stale if an attacker later modifies the resource. As an example, assuming resource X is decomposed into four tasks (X.[1-4]). If tasks X.1, X.2, X.3 are performed and then the attacker compromises resource X region X.2, the X.4 measurement will complete and resource X will be declared unchanged. The compromise will only be discovered at the next measurement of X.2.

Scrubbing attacks also are challenges for SMM-based runtime integrity mechanisms [3]. In these attacks, malware anticipates when an integrity measurement will be occurring and hides its traces before the measurement. After the measurement completes, malware can restore itself. While not a complete remedy, EPA-RIMM's time-sliced measurement scheduling adds a complication to the work of an attacker relying upon transient attacks as it is not clear when inspections will be made and which resource will be examined at a given point in time.

It is also difficult for SMM RIMMs to be completely stealthy. A motivated attacker could leverage timing information to ascertain losses of control. The developers of a stealthy SMM-based debugger, MALT, note that while they were able to adjust various system timers in SMM to hide their operation, a dedicated attacker could send an encrypted message to a remote timing server to get accurate sense of time [28]. For these reasons, while SMM RIMMs operate independently from host software, complete stealth appears infeasible.

## 3   EPA-RIMM Minnowboard Prototype

We developed a prototype based on the Minnowboard Turbot system [17]. The prototype implements a subset of the EPA-RIMM architecture. Our prototype system does not have an out-of-band communication mechanism, therefore we demonstrate the functionality using an in-band mechanism. This is sufficient for the research prototype since it does not share the security requirements of a production system. We implemented four separate modules: The Backend Manager (BEM), an in-band Host Communications Manager (consisting of the "Frontend Manager" (FEM) and "Ring 0 Manager" (R0M) modules), and the Inspector. The BEM runs on a network server while the other components reside on the monitored system. Our current prototype does not implement the Diagnosis Manager.

**Hardware**. We selected the Minnowboard as it is unique in offering allowing the ability to recompile and re-flash its firmware and supports the x86 CPU architecture. This board, however, has less computational resources than a typical server platform that EPA-RIMM would be running on. The Turbot features a dual core Intel Atom e3826 processor with a base clock of 1.46 GHz, 2 GB RAM, onboard gigabit Ethernet, and supports a SATA disk [26]. We installed Ubuntu 12.04 64bit with a 3.7.16 kernel on the attached SSD.

**Firmware**. We modified the SMI handler in the UEFI source (Minnowboard Max firmware version 0.94) by creating an EPA-

RIMM DXE_SMM_DRIVER which registers a new software SMI using the EFI_SMM_SW_DISPATCH_PROTOCOL. We integrated OpenSSL 1.0.2d support into the Inspector for SHA256 hashing for measurements, AES256-CBC for encryption, and HMAC SHA256.

**Backend Manager (BEM)**. We implemented our Backend Manager in Python. It sets up a network socket connection with a set of Frontend Managers (one per monitored system). We add a set of tasks directly to the BEM's priority queues that it maintains for each Monitored Node. As each monitored system may process tasks at different rates, the number of tasks that fit in a given Bin size may vary across systems.

**Frontend Manager (FEM)**. The FEM receives the signed, and encrypted Bin from the BEM and writes it to a /proc interface that is registered by the Ring 0 Manager. After each Bin has been processed, the Frontend Manager retrieves the results from the /proc interface and send them to the BEM.

**Ring 0 Manager**. The Ring 0 Manager (Linux kernel module) receives the Bin from the Frontend Manager via the write to the /proc interface. The Ring 0 Manager then records the virtual memory location of the Bin in a CPU register. After this, it triggers the measurement SMI by writing a pre-arranged value to port 0xB2.

**Inspector**. For the provisioning phase, the Inspector computes hash values for the specified operation (MSR, CPU register, or memory region) and writes the hash value back into the results data structure. For subsequent measurements, the Inspector receives an initial hash value from the BEM for comparison.

### 3.1 Attack Detection Using the Prototype

To demonstrate EPA-RIMM's ability to detect rootkits, we implemented and ran four known compromise techniques with a Linux-based platform:  1. changes to the Interrupt Descriptor Table Register (IDTR); 2. changes to CR4.SMEP; and 3. a change in a Linux kernel code section, as would be done in a kernel code injection rootkit. 4. Hooking of the system call table via a write to CR0 and modification of kernel read only data structures.

*3.1.1 Interrupt Descriptor Table Register (IDTR)*.  The IDT hooking rootkit attack targets the Interrupt Descriptor Table Register (IDTR) which holds the virtual address of the Interrupt Descriptor Table (IDT) [20]. The IDT associates interrupts or exceptions with their handler routines as IDT Entries. There are two variants of this attack: insertion of malicious code into one or more IDT entries; or overwriting of the IDTR value to point to a new, modified copy of the entire IDT. We demonstrate the EPA-RIMM prototype's detection of a change in the IDTR.

*EPA-RIMM Check: {Command = Register (IDTR) }*

*Normal Functioning*

1.  *Provisioning:* The IDTR Check is enabled and the BEM adds it as a single Task in the priority queue.

2. *Scheduling:* The BEM schedules the IDTR Task by placing it in a Bin and sending it to the FEM.

3. *Frontend Processing:* The FEM receives the Bin and writes it to the Ring 0 Manager /proc interface. The Ring 0 Manager generates a measurement SMI providing the physical address of the Bin.

4. *Inspector:* The Inspector receives the SMI; it uses a known CPU register value to determine the Bin address, and handles each Task in the Bin sequentially. In this case, the Inspector decrypts the Bin, checks its signature, and reads the upper and lower IDTR fields from the SMRAM Save State map for each CPU thread, records a hash for the measurement in the results data structure, signs and encrypts the results.

*Compromise:* Once the Inspector has provisioned the initial values, we performed an attack on the IDTR by changing its value from kernel code, mimicking the operations of a rootkit. This attack uses a published example of saving the original IDT, copying it to a newly allocated kernel page, saving the memory address of the newly allocated kernel page, and loading the address of the new page into the IDTR register [18].

*Detection:* At the next IDTR measurement, the Inspector re-checked the IDTR value from the SMRAM Save State Map and flagged the difference in the results that it returns.

### 3.1.2. CR4.SMEP Disable.

One important CPU-based security feature is Supervisor Mode Execution Protection (SMEP). This prevents ring 0 code from being able to execute instructions from user-mode pages. The SMEP feature is enabled by setting bit 20 in the CR4 control register.

*EPA-RIMM Check: {Command = Register (CR4) }*

*Normal Functioning*

1. *Provisioning:* We prepare for an initial measurement of the CR4 value by enabling the CR4 register Check.

2. The flow proceeds similarly to the previously described steps 2-4 in Section 3.1.1 with the difference that the Inspector looks up the CR4 register values for both CPU threads in the respective SMRAM Save State Maps.

*Compromise.* We disabled CR4.SMEP using a kernel module utility.

*Detection.* At the next measurement of the CR4 value, the Inspector detected the change on both CPU threads due to a hash difference.

### 3.1.3. Kernel Rootkit Code Injection.

The Snakso rootkit appeared in 2012 and targeted 64-bit Linux kernels [19]. After becoming entrenched in the Linux kernel, it replaced the kernel's `tcp_sendmsg` function with a compromised version that injected malicious web browser iFrames into HTTP traffic. This type of rootkit could be detectable by hashing portions of the Linux kernel as the rootkit modified or replaced the `vfs_readdir`, `vfs_read`,

`filldir64`, and `filldir` kernel functions. For simplicity, we modified kernel function `ftrace_raw_event_xen_mc_entry`, that was not utilized during our system's execution.

*EPA-RIMM Check: {Command = Virt Memory (Kernel Code Start to Kernel Code End), Length = 4KB }*

*Normal Functioning*

1. Provisioning: EPA-RIMM is initialized with 4K hash values for the kernel code memory region upon the first measurement.

2. The flow proceeds similar to the previously described steps 2-4 in Section 3.1.1.

3. The Inspector receives the Bin and performs a SHA 256 hash for the specified address and length. The Inspector returns the hash result to the BEM.

*Compromise.* To simulate a memory compromise, we overwrote an area of kernel code using a debug capability.

*Detection.* Upon the subsequent re-measurement, the Inspector identified a difference in the current and provisioned hashes for the task covering the function's code and returns a changed result.

### 3.1.3 System Call Hooking.

One common attack strategy for rootkits is to hook critical kernel data structures such as the system call table. This attack endeavors to trigger attacker-provided code to execute when a trigger is invoked. An example of this attack is a demonstration attack published by *forb1dd3n* called *sys_call_hijack* [50]. This attack searches for the location of the system call table in the System.map file, clears the write-protect bit in CR0, modifies the system call table to insert a new attacker-provided function, and re-enables the write-protect bit in CR0. EPA-RIMM has two potential Checks that can detect this attack, kernel read-only data structure Check and a CR0 Check. As the attack would be intended to persist, the kernel read-only data structure Check would be able to detect the change upon its next execution as the attack would necessitate a change in the system call table. It is also theoretically possible that EPA-RIMM could detect a change in progress for the CR0 register, however, EPA-RIMM would need to be invoked at precisely the time where the CR0 register had changed (to disable write-protect) and later returned to its initial value.

*EPA-RIMM Check 1: {Command = Virt Memory {Kernel Read Only Data Structure Start to Kernel Read Only Data Structure End}*

*EPA-RIMM Check 2: {Command = Register (CR0) }*

*Normal Functioning*

1. Provisioning: EPA-RIMM is initialized with hash values for the kernel read only data sections and also the CR0 value hashed. The flow proceeds similarly to steps 2-4 in Section 3.1.1.

*Compromise:* We load the sys_call_hijack rootkit and monitor EPA-RIMMs alerts.

*Detection:* The Inspector identifies a change in the kernel read-only data structures. Detecting a change in the CR0 register would require EPA-RIMM execution at the time that it had changed.

## 3.2 Discussion

In this section, we demonstrated the ability of EPA-RIMM to detect rootkit techniques that compromised the IDTR, CR4 register, kernel code, and read-only kernel data structures. The prototype's extensible design allows dynamically changing the set of measurements performed by EPA-RIMM by creating and scheduling new tasks to be performed by the Inspector. The prototype's ability to process multiple tasks of smaller durations demonstrates that larger measurement tasks such as those covering the kernel memory region can be accomplished in smaller portions reducing the impact on the running system.

## 4. Performance Evaluation

In this section, we evaluate the performance of EPA-RIMM and provide a simple performance model for EPA-RIMM overheads.

## 4.1 A simple performance model

The total time for an SMI-based measurement can be represented as $T_m$, where

$$T_m = T_{entry} + T_{work} + T_{exit}$$

and $T_{entry}$ is the time to enter SMM, $T_{work}$ is the total time to accomplish the measurement, and $T_{exit}$ is the time to transition out of SMM.

$T_{work}$ is platform dependent and varies with the specific type of measurement. $T_{work}$ can be broken down as follows:

$T_{work} = T_{verify} + T_{task} + T_{eval} + T_{package} + T_{copy}$
$T_{verify} = T_{decrypt} + T_{sigcheck}$
$T_{package} = + T_{sign} + T_{integrity} + T_{encrypt}$
(see **Table 5**)

**Table 5 EPA-RIMM Latency Analysis**

| SMI Latency Component | Cost dependent on: |
|---|---|
| $T_{entry}$ : Transition time to enter SMM | Platform, CPU C-state, firmware |
| $T_{decrypt}$ : Bin Decryption | Number of tasks, decryption algorithm performance |
| $T_{sigcheck}$ : Bin signature check | Signature verification algorithm |
| $T_{task}$ : Task Processing | Number and type of tasks in Bin |
| $T_{eval}$ : Measurement evaluation | Number of tasks in Bin |
| $T_{sign}$ : Creation of results signature | Signature creation algorithm |
| $T_{encrypt}$ : Results encryption | Number of results, encryption algorithm performance |
| $T_{integrity}$ : Integrity measurement (e.g. HMAC) of bin or results | MAC performance and size of bin or results |
| $T_{copy}$ : Copy of results to HCM | Number of results |
| $T_{exit}$ : Return from SMM | Platform, firmware version |

## 4.2 Latency Guidelines

To get a starting baseline for SMI latency on the Minnowboard Turbot, we measured the cost for a basic software SMI by taking CPU timestamps before and after a sequence of 1000 basic SMIs that wrote a value of 0 to port 0xB2. For this testing, we set the CPU to run at the C0 C-state to avoid the variable performance impact of C-state transitions. CPU C-states allow power savings when the CPU is idle and transitioning between states can take varying amounts of time depending on the C-state transition [51]. The generated SMI has minimal processing and the average amount of time to process each SMI was 105 microseconds. Meeting the SMI latency target of $T_m$ = 150 *u*sec thus requires $T_{work} <= 45$ *u*sec.

In Figure 2a we show the minimal cost for the CR4 measurement used in Section 3, which includes access time to CR4 values from the SMRAM Save State Map for both CPU threads and comparison with the expected values. The average time was 0.6 microseconds. This measurement consists of items $T_{task} + T_{eval} + T_{sign}$ in the performance model.

The kernel memory hash costs ($T_{task}$) vary by the amount of memory hashed (Figure 2b). We observe that the hash costs show low variance due to the lack of interruption in SMM. In Figure 2c, we show the total Bin costs $T_m$ for kernel memory hashing. These Bin costs assume one Task per Bin and each bar consists of all EPA-RIMM SMI overheads enumerated in the performance model. SMI transition time consists of the time entering and exiting SMM, $T_{entry} + T_{exit}$. The non-hash Inspector costs consist of all Inspector overhead excluding the hash measurement operation: $T_{decrypt} + T_{sigcheck} + T_{eval} + T_{sign} + T_{encrypt} + T_{copy}$. (Note: $T_{integrity}$ is not yet included in these numbers.) These costs are essentially constant. In contrast, the hash costs are completely dependent on the chosen hash size. We have not yet applied performance optimizations for the Inspector so it is possible that the Inspector costs can be reduced further.

Given the tight SMI processing timeframes, care needs to be taken in ensuring that signature checking, hashing, and encryption times are kept minimal. Presently, our prototype Inspector performs a simple signature check that is not RSA-based which reduces the time requirements. A costlier signature check would increase EPA-RIMM costs for items $T_{sigcheck}$ and $T_{sign}$.

In **Table 6**, we compare EPA-RIMM's SMI durations and frequency compared to the other SMM RIMMs. We observe that due to Check decomposition, EPA-RIMM is able to significantly undercut the times for previous methods, coming close to the SMI latency expectations on very modest CPU hardware. By limiting SMI durations, negative interactions with device drivers, application performance, and host software code can be greatly reduced [2].

## 4.3 Performance Impact

To evaluate the performance impact of EPA-RIMM, we compared three scenarios: A. No SMIs (Baseline), B. Frequent 8KB hashes (22 hashes of 8KB size every second which was the fastest rate supported by our prototype), C. Infrequent large hashes (two hashes of 1MB size every second). We utilized the Phoronix Memcached, SQLite, and cachebench benchmarks and compared the throughput achieved by both EPA-RIMM scenarios against our baseline in which no SMIs occurred [45]. The first two workloads are representative of server workloads that could be running on EPA-RIMM monitored servers and the cachebench workload illustrates CPU cache-related effects. For Memcached we utilized the 'get' benchmark which retrieves values from the memory object cache. SQLite exercises the disk subsystem. For cachebench, we focused on cache read performance, however, cache writes and read-modify-update performed very similarly.

Our test system utilized the in-band Ring 0 Manager and Frontend Manager which consumed a small percentage of CPU time. Production EPA-RIMM implementations using the specified out-of-band interface would not have these costs.

For the 22 x 8KB hash EPA-RIMM scenario, shown in Figure 2d, we observe that CPU cache results are only slightly impacted as the workload still achieves 99.23% of its baseline throughput. Memcached shows a larger impact as the scenario achieves 94.77% of its baseline throughput. Similarly, SQLite achieved 96% of the baseline performance. The 2x1MB hash scenario performance are similar at 94.3%, 94.09%, and 95.78% respectively. Beyond the amount of CPU time taken away, the cachebench results indicate that there is an additional impact on the CPU caches greater than the time spent in SMM. This may be a combination of host-side Ring 0 Manager, EPA-Frontend network traffic, and Inspector activities. As the Backend Manager can readily control the frequency of measurements as well as the measurement size, EPA-RIMM can dial up or down the system impact via these two factors.

#### Table 6 SMI Time Comparison of SMM RIMMs

| SMM RIMM | SMI Duration | Frequency |
|---|---|---|
| HyperCheck | 40ms | 1 per second |
| HyperSentry | 35ms* | 1 per 8 or 16 sec |
| SPECTRE | 5 to 32ms | 16 per second to 1 per 5 |

| | | seconds |
|---|---|---|
| EPA-RIMM Minnowboard | 0.24 ms+ | Dynamic |
| (Goal) | 0.15 ms | Not specified |

*HyperSentry utilizes several SMIs to complete the measurement, totaling 35 ms.

#### Table 7 Backend Manager SHA256 Memory Storage Space Requirements

| Hash Size (KB) | # hashes | Aggregate SMM Time (uSecs) | BEM Space Req. (Bytes) for SHA 256 |
|---|---|---|---|
| 0.5 | 18,432 | 4,442,112 | 589,824 |
| 1 | 9,216 | 2,359,296 | 294,912 |
| 2 | 4,608 | 1,331,712 | 147,456 |
| 4 | 2,304 | 827,136 | 73,728 |
| 8 | 1,152 | 569,088 | 36,864 |

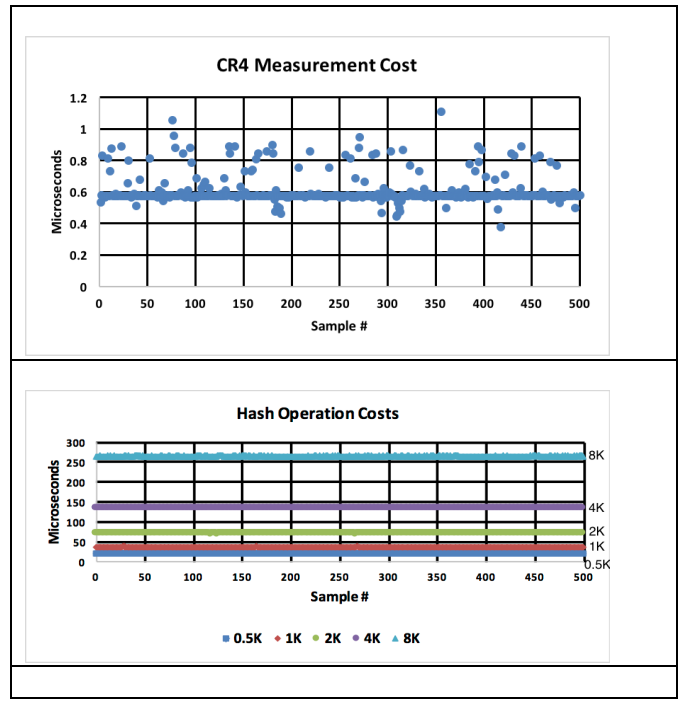

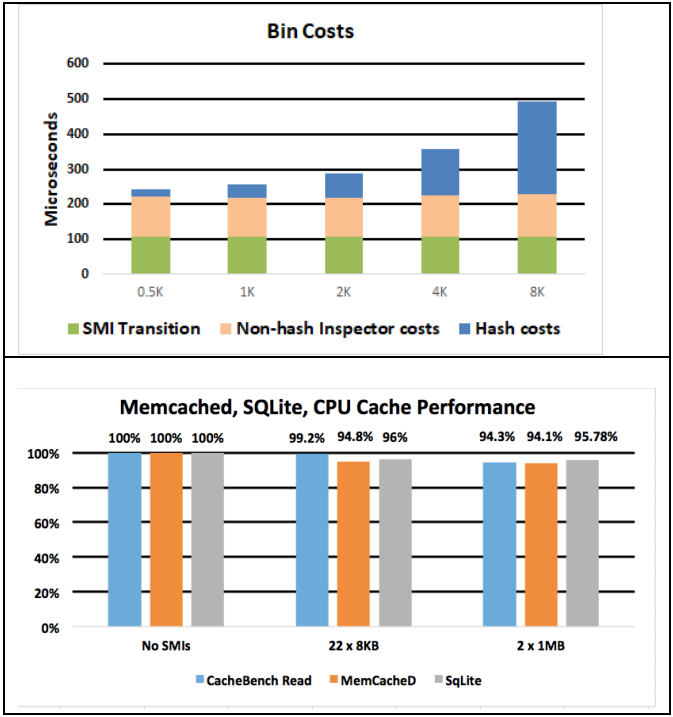

**Figure 2 (top to bottom) (a) Performance cost of CR4 Measurement, (b) SHA256 Hash Times (SMM Inspector), (c) Total EPA-RIMM SMI durations for varied hash sizes, (d) Application and Microbenchmark Performance for cachebench read, Memcached, and SQLite.**

## 4.4 Check Decomposition Trade-offs

Check decomposition allows EPA-RIMM to minimize the amount of time spent in a single SMM measurement session. However, it involves an overall efficiency trade-off as smaller amounts of work processed per SMI result in more SMIs in total.

*Methodology*. We gathered a list of Linux kernel code sections utilizing the System.map file for the 3.7.6 kernel running on our monitored node. The size was around 9 MB. For these sections, we examined decompositions of 1K, 2K, 4K, 8K hash sizes. We calculate the wall-clock time to drain the queue with the Backend Manager scheduling 22 bins per second. We utilized the total Bin processing time as measured by the Ring 0 Manager in which we gathered the CPU's time-stamp counter before and after the measurement SMI trigger. We also counted the number of SMIs by logging MSR_SMI_COUNT (0x34) in the Ring 0 Manager to verify the rate of SMIs. We note that here we assume a fixed rate of SMIs, but production EPA-RIMM instances would randomize the Bin scheduling to be less predictable.

*Results*. With a fixed rate of issuing the bins and the larger hash size, the 8KB queue takes the least amount of time at 52.4 seconds to hash the kernel code sections. Each 8KB measurement, requires 494 microseconds in SMM. As there would be 1,152 hashes over 8KB memory regions, this would require 36,864 bytes of hash storage space on the Backend Manager when using SHA256. On the other end of the spectrum, a queue of 0.5 KB hashes would be accomplished in 838 seconds, requiring 241 microseconds in SMM, and 589,824 bytes of hash storage on the Backend Manager. Figure 3 and **Table 7** provide results for other measurement hash sizes.

The third column in **Table 7** shows the aggregate amount of time in SMM across the necessary number of SMI sessions required to complete the kernel memory hash. For example, in the most extreme case, the 0.5 KB hash requires an aggregate of 4,442,112 microseconds in SMM to complete the kernel hash as there are 18,432 SMIs required (assuming one Task per Bin), while the 8KB hash can perform the same work in 569,088 microseconds and only require 1,152 SMIs. The appropriate trade-off between SMI latency and overall measurement throughput on a given system, are tunable parameters: Bin size and frequency. These knobs can be adjusted upwards during the spread of an attack or dialed down to emphasize application performance.

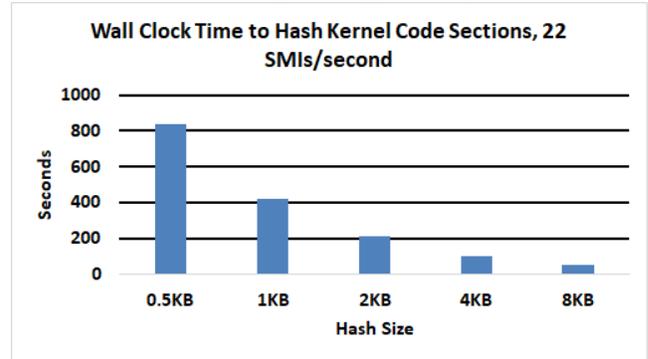

**Figure 3 Hash Time**

## 4.5 Encryption and Hashing Performance

As the Inspector SMI latencies would benefit from a performance increase in encryption, decryption, and hashing, we compared the hashing and encryption performance on our prototype system with a higher-end laptop CPU. For this study, we utilized OpenSSL 1.1.0b. We prepared a Minnowboard Turbot with an Atom E3826 with a base clock of 1.46 GHz and also an Intel NUC utilizing a laptop-class CPU (Core i5 6260U CPU) with a base clock of 1.86 GHz. We measured 1K block sizes for AES 256 CBC encryption and SHA 256 performance without AESNI acceleration. The Intel Core i5 6260U performed 4.41x better than the Intel Atom E3826 on AES 256 CBC. For SHA256, the Core i5 achieved 5.26x greater throughput than the Atom for this workload. These results show that our prototype could reduce time spent in $T_{decrypt}$, $T_{task}$ (hashing), $T_{encrypt}$ with a higher performance CPU.

## 4.6 Performance Conclusions

Our performance results show that EPA-RIMM comes very close to the SMI latency expectations. This result on modest CPU hardware is orders of magnitude better than other SMM RIMMs, demonstrating the effectiveness of the decomposition approach. We expect that higher performing CPUs could improve upon these numbers due to reduced costs for hashing, decryption, and encryption operations. Factors that could increase the time spent in SMM consist of HMAC and a more robust signature checking mechanism. While decomposition avoids prolonged SMM sessions, there is a trade-off in overall measurement efficiency. Finally, the memory requirements on the Backend Manager are reasonable as a single Backend Manager could support many thousands of monitored nodes and multiple Backend Managers can be readily deployed in the EPA-RIMM framework as necessary.

## 5 Security Analysis of EPA-RIMM

In this section, we describe our assumptions and analyze potential threats against EPA-RIMM. We consider attacks against SMM and EPA-RIMM components and on requests and results, side channels, initial measurements, infrastructure compromise and denial of service, crypto/signing attacks, and transient evasion attacks.

*Assumptions*. We assume that initial measurements can be gathered during a provisioning process. We also assume that SMRAM is well-protected and leverages available hardware protections including SPI protections over the BIOS chip and proper SMRR configuration. The CHIPSEC tool can be used to verify proper platform SMM configuration [23]. EPA-RIMM also assumes the presence of an out-of-band network interface to allow communication of measurements requests and results. We assume the out-of-band interface is not malicious.

*Threat Model*. EPA-RIMM targets scenarios where an attacker has gained control over the operating system or hypervisor at runtime. This can include code injection into these privileged layers.

*Inspector*. The Inspector, residing within SMM, may be targeted by the attacker. One potential attack is a confused deputy attack in which the attacker attempts to trick the Inspector into overwriting SMRAM memory or other privileged memory. The Inspector should be written such that it checks any input buffers to ensure that they don't reside within the SMRR. The Inspector should also communicate directly to the Host Communications Manager and not write data into OS-controlled memory. Attacks on EPA-RIMM's Inspector could attempt to forge a measurement request from a malicious Backend Manager in order to gain additional insights into operation of the system. For this reason, it is important that the Inspector and Backend Manager properly authenticate communications with each other using a signature verification over the communications. The Inspector and Backend Manager also use encryption for their communications to prevent eavesdropping or tampering. The use of a nonce prevents replay attacks in which previous measurements are passed off as current measurements. The Inspector should not return more information than is required to determine an unwanted change has occurred. By returning hashes instead of actual measurement values, the Inspector helps limit its potential usefulness as a side channel.

*Initial Measurements and EPA-RIMM launch*. The treatment of initial measurements requires special handling. They should be provisioned by an Administrator in an offline environment to avoid compromised values appearing as pristine values. In homogeneous environments with a large number of systems running identical kernel and OS versions, it may be possible to gather an initial measurement on one or more nodes and store their results for comparison against other identically-configured nodes. New kernel versions would require re-provisioning due to differences in memory layout. Once the host software launches with a trusted boot, EPA-RIMM can begin servicing measurement requests over the out-of-band Host Communications Manager interface.

*Infrastructure Compromise and Denial of Service*. If a Denial of Service were to affect the DM or Backend Manager, the flow of measurement requests would slow or cease. Monitoring of the flow of EPA-RIMM measurements would be necessary to identify this type of attack. If the Diagnosis Manager were to be compromised, it would be possible to misdirect EPA-RIMM to monitor an unrelated set of resources while an attack executes or share false reports of attack detection in an attempt to guide measurements on other nodes. For this reason, the EPA-RIMM Administrator should monitor and investigate the threat intelligence exchanged by EPA-RIMM and also audit the Checks that are being performed for unexpected changes. A compromise of the Backend Manager could expose the hash database. However, this is of limited use as the Backend Manager only stores hashes as opposed to memory contents.

*Transient Evasion Techniques*. All snapshot-based periodic inspections have the potential to miss attack detection if signs of the attack were not present at the measurement interval. EPA-RIMM, unlike other SMM RIMMs, can be used to measure more frequently, in smaller portions to reduce the amount of time between periodic measurements. Additionally, varying the set of measured items dynamically at runtime leads to less predictable measurements which complicates the attacker's task.

*Host-side Memory Visibility*. A recent Minnowboard Max firmware release (version 0.97) has added enhanced protections for SMM code that would require modifications for EPA-RIMM to function [24]. These changes unmap host-side memory from the SMM code and also check to see if host-side memory accesses are within the Communications Buffer (CommBuffer). As EPA-RIMM relies upon the ability to inspect host memory, modifications to allow host memory to be readable would be necessary for functionality. We have prototyped changes that allow proper EPA-RIMM functioning on the 0.97 codebase and are investigating enhancements to EPA-RIMM to find a balance between efforts to constrain SMM memory and EPA-RIMM's measurement requirements.

## 6 RELATED WORK

Rootkit detection could in theory be rendered unnecessary by accurate and complete rootkit prevention mechanisms (see SecVisor [13] Lares [36] PerspicuOS [35] and Microsoft Virtual Secure Mode (VSM) [38]). In the absence of perfect instantaneous prevention, these are complementary to EPA-RIMM.

Development of rootkit detection mechanisms has been characterized as a "race to the bottom" in which more privileged layers are examined when threats appear [43]. Pure software-based approaches such as rkscan [8][9] and St. Michael [29] can be easily deployed; however, effective protection of the measurement mechanism is infeasible because it runs in the same context as the malicious code it is trying to detect. Hardware-based mechanisms can achieve a greater isolation from malicious code due to lower-level hardware protections over their agent. Due to their isolated contexts, hardware peripheral-based approaches such as Copilot [10], MGUARD [11], Vigilare [6], and DeepWatch [12] are limited in their ability to gain full access to the necessary CPU and memory state. MGUARD has the unique capability to detect SMM rootkits due to its integration into the memory module itself. Vigilare is able to catch transient attacks, which compromise the system then hide traces of their presence in order to avoid detection. However, as Vigilare lacks access to the CPU registers, it is vulnerable to relocation attacks. DeepWatch can detect and remediate low-level virtualization-based rootkits by using chipset hardware out of band of the host CPU. Flicker [16] and Sailer, Zhang et al. [15] leverage the Trusted Platform Module (TPM) to dynamically measure executable content such as applications, the kernel, and kernel modules on a running Linux system before invocation or use [33][44]. McAfee Deep Defender [14] utilizes a security hypervisor to monitor several key Windows resources in a virtualized Windows guest. These resources include the IDT, SSDT, DKOM list, kernel code sections, certain device drivers, among others [30]. This approach benefits from strong isolation between the monitored environment and the security hypervisor and can also provide quick detection of improper accesses as they occur, but at a significant performance cost [31]. Work by Demme et al. leverages CPU performance counters to detect anomalous rootkit behavior [32]. This approach can be integrated into EPA-RIMM as an additional type of Check, and is complementary to our efforts.

Previously proposed SMM RIMMs include HyperSentry [3], HyperCheck [4], and SPECTRE [5]. Our work is influenced by these prior efforts and also extends the state of the art, as follows:

**Context.** Our approach is influenced by SPECTRE's methods for understanding host-side data structures, however we do not build this understanding into the Inspector, instead, allowing the Check contents to direct the Inspector's operation [5].

***SMM RIMM Measurement Trigger.*** The EPA-RIMM architecture utilizes a similar mechanism to HyperSentry to invoke SMM-RIMM measurements out-of-band; This approach is different from HyperCheck's network card approach or SPECTRE's chipset timer; the former is vulnerable to network card firmware-based attacks, while the latter is limited to a small number of scheduling options supported by chipset hardware and thus would be relatively predictable.

***Location of measurement code/Extensibility.*** While EPA-RIMM's Inspector resides in SMM, it does not build in hostside context (as SPECTRE and HyperCheck do), instead processing measurement requests. When the OS/VMM is updated, the measurement requests could change but the Inspector code does not require an update as it supports fundamental building blocks for new measurements. This significantly improves the ability to react to emerging threats and to thwart efforts of rootkits to avoid detection by the monitoring tool.

***Hardware Requirements.*** HyperSentry uses specific SMI chipset routing and a modified BMC. HyperCheck utilizes an SMM-enabled network card, SPECTRE utilizes a dedicated network card for SMM communication. EPA-RIMM can utilize any out-of-band mechanism such as a BMC.

***Measurement Decomposition.*** EPA-RIMM is unique in natively supporting the ability to decompose long-running measurements. This allows it to avoid prolonged SMI delays that would result in system perturbation [2]. The approach complicates an attacker's understanding of what is being monitored as EPA-RIMM interleaves decomposed measurements which results in additional non-determinism.

# 7  CONCLUSIONS

In this paper we have introduced EPA-RIMM, a Runtime Integrity Measurement Mechanism that presents a flexible and scalable way to utilize System Management Mode for rootkit detection. EPA-RIMM represents a significant re-design of the SMM RIMM concept, providing several new contributions. We utilize Check decomposition to bound the perturbation and enable a wide variety of measurements to be performed over periods of time, ensuring scalability when new or more complex measurements are required; thus rendering continuous monitoring feasible. This also introduces a degree of nondeterminism, as smaller tasks are interleaved and scheduled at randomized intervals, thus reducing predictability. EPA-RIMM's Check description API allows crafting of new Checks to respond to emerging threats in the environment. We have also presented a Minnowboard-based EPA-RIMM prototype that can be used for research and education purposes. We plan an open source release of the prototype code.

The EPA-RIMM project is ongoing. Our current focus spans several important thrusts. We are continuing development of the Diagnosis Manager, exploring a number of new Checks and types of measurements, particularly focused on the hypervisor and on automated Check triggers. We are developing a simulator to evaluate scheduling. Finally, one key performance impact of SMM time is that all host CPUs are unavailable for application processing while they are in SMM, and for security we do not want to allow them to run application code while we are performing our measurements. To address this scaling issue, we are developing the needed infrastructure to utilize all cores for SMM-mode measurements.

## ACKNOWLEDGMENTS

This work was supported in part by NSF Award #1528185. The authors gratefully acknowledge the contributions of the EPA-RIMM team including Tejaswini Vibhute, John Fastabend, Cody Shepherd.

## REFERENCES

[1] Mannthey, K, "System Management Interrupt Free Hardware," IBM Linux Technology Center. http://linuxplumbersconf.org/2009/slides/Keith-Mannthey-SMIplumers-2009.pdf, accessed July 31, 2013.

[2] Delgado, B., Karavanic. K. "Performance Implications of Systems Management Mode", IISWC, Portland, OR, 2013.

[3] Azab, A., Ning, P., Wang, Z., Jiang, X., Zhang X., Skalsky, N. "HyperSentry: enabling stealthy in-context measurement of hypervisor integrity," CCS. Chicago, IL, 2010.

[4] Wang, J., Stavrou, A., Ghosh, A. "HyperCheck: a hardware-assisted integrity monitor," Lect. Notes Comput. Sci. Lecture Notes in Computer Science (including subseries Lecture Notes in Artificial Intelligence and Lecture Notes in Bioinformatics) 6307 LNCS: 158-177, 2010.

[5] Zhang, F., Leach, K. Sun, K., Stavrou, A. "SPECTRE: A Dependable Introspection Framework via System Management Mode", 43rd Annual IEEE/IFIP International Conference on Dependable Systems and Network (DSN), Budapest, Hungary, 2013.

[6] Moon, H., Lee, H., Kim, K., Paek, Y., Kang, B.B. "Vigilare: toward snoop-based kernel integrity monitor", Proceedings of the 2012 ACM Conference on Computer and Communications Security, New York, NY, 2012.

[7] Triplett, J and Triplett, B, "BITS: BIOS Implementation Test Suite,"http://www.linuxplumbersconf.org/2011/ocw/system/presentations/867/original/bits.pdf

[8] http://www.hsc.fr/ressources/outils/rkscan/

[9] http://seclists.org/incidents/2000/Oct/165

[10] Petroni, N., Fraser, T. Molina, J., Arbaugh, W. "Copilot – a coprocessor-based kernel runtime integrity monitor", Proceedings of the 13th USENIX Security Symposium. San Diego, California, 2004.

[11] Liu, J., Lee, J., Zeng, J., Wen, Y., Lin, Z., Shi, W. "CPU Transparent Protection of OS Kernel and Hypervisor Integrity with Programmable DRAM", ISCA, Tel Aviv, Israel, 2013.

[12] Bulygin, Y., Samyde, D. "Chipset-based approach to detect virtualization malware a.k.a. DeepWatch", Blackhat USA, 2008.

[13] Seshadri, A., Luk, M.,, Qu, N., Perrig, A. "SecVisor: A Tiny Hypervisor to Provide Lifetime Kernel Code Integrity for Commodity OSes", SOSP '17, Stevenson, Washington, 2007.

[14] http://www.businesswire.com/news/home/20111018006406/en/McAfee-Deep-Defender-Helps-Reinvent-Industry-Approach#.VP_Tn0Lwwg M

[15] Sailer, R., Zhang, X., Jaeger, T., van Doorn, L. "Design and Measurement of a TCG-Based Integrity Measurement Architecture", Proceedings of the 13th USENIX Security Symposium. San Diego, California, 2004.

[16] McCune, J., Parno, B., Perrig, A., Reiter, M., Isozaki, H. "Flicker: An Execution Infrastructure for TCB Minimization", EuroSys '08, Glasgow, Scotland, 2008.

[17] http://www.minnowboard.org/

[18] https://github.com/RichardUSTC/intercept-page-fault-handler

[19] http://securelist.com/blog/incidents/34623/new-64-bit-linux-rootkit-doing-iframe-injections-30. Accessed March 24, 2015.

[20] http://phrack.org/issues/59/4.html, accessed November, 30, 2015

[21] Kallenberg, C., Kovah, X., "BIOS Necromancy: Utilizing "Dead Code" for BIOS Attacks", HITB 2015, Singapore, 2015.

[22] Wojczuk, R., Tereshkin, A., "Attacking Intel BIOS", Black Hat USA 2009.

[23] Loucaides, J., Bulygin, Y. "Platform Security Assessment with CHIPSEC", CanSecWest, Vancouver BC, Canada, 2014.

[24] https://firmware.intel.com/projects/minnowboard-max

[25] https://software.intel.com/en-us/articles/intel-sha-extensions

[26] http://ark.intel.com/products/78477/Intel-Atom-Processor-E3826-1M-Cache-1_46-GHz

[27] http://ark.intel.com/products/91160/Intel-Core-i5-6260U-Processor-4M-Cache-up-to-2_90-GHz?q=6260U

[28] Zhang, F., Leach, K., Stavrou, A., Wang, H., Sun, K. "Using Hardware Features for Increased Debugging Transparency", IEEE S&P, 2015.

[29] https://github.com/tomasz-janiczek/stmichael-lkm

[30] McAfee Deep Defender Data Sheet, McAfee, http://www.mcafee.com/us/resources/data-sheets/ds-deep-defender.pdf

[31] McAfee Deep Defender Technical Evaluation and Best Practices Guide, Version 1.0, McAfee. https://kc.mcafee.com/resources/sites/MCAFEE/content/live/PRODUCT_DOCUMENTATION/23000/PD23874/en_US/Deep_Defender_Best_Practices_Guide_Aug_2012.pdf

[32] Demme, J., Maycock, M., Schmitz, J., Tang, A., Waksman, A., Sethumadhavan , S., Stolfo, S. "On the Feasibility of Online Malware Detection with Performance Counters", ISCA 2013, Tel Aviv, Israel, 2013.

[33] Sailer, R., Zhang, X., Jaeger, T., van Doorn, Leendart. "Design and Measurement of a TCG-Based Integrity Measurement Architecture", Proceedings of the 13th USENIX Security Symposium. San Diego, California, 2004.

[34] Futak, A., Bulygin, Y, Bazhaniuk, O., Loucaides, J., Matrasov, A., Gorobets, Mikhail, "BIOS and Secure Boot Attacks Uncovered", Eko Party 10,
http://www.intelsecurity.com/resources/pr-bios-secure-boot-attacks-uncovered.pdf , 2014. Accessed February 1, 2017.

[35] Dautenhahn, N., Kasampalis, T., Dietz, W., Criswell, J., Adve, V. "Nested Kernel: An Operating System Architecture for Intra-Kernel Privilege Separation", ASPLOS '15, Turkey, 2015.

[36] Payne, B., Carbone, M., Sharif, M., Lee, W. "Lares: An architecture for Secure Active Monitoring Using Virtualization", IEEE Symposium on Security and Privacy, Oakland, California, 2008.

[37] Song, C., Lee, B., Lu, K., Harris, W., Kim, T., Lee, W. "Enforcing Kernel Security Invariants with Data Flow Integrity", NDSS '16, San Diego, California, 2016.

[38] https://blogs.technet.microsoft.com/ash/2016/03/02/windows-10-device-guard-and-credential-guard-demystified/, accessed March 12, 2017.

[39] Hund, R., Willems, C., Holz, T. "Practical Timing Side Channel Attacks Against Kernel Space ASLR", IEEE Symposium on Security and Privacy '13, San Francisco, CA, 2013.

[40] Gruss, D., Maurice, C., Fogh, A., Lipp, M., Mangard, S. "Prefetch Side-Channel Attacks: Bypassing SMAP and Kernel ASLR", CCS '16, Austria, 2016.

[41] Jang, Y., Lee, S., Kim, T. "Breaking Kernel Address Space Randomization with Intel TSX", CCS '16, Austria, 2016.

[42] Evtyushkin, D., Ponomarev, D., Abu-Ghazaleh, N. "Jump over ASLR: Attacking Branch Predictors to Bypass ASLR", IEEE International Symposium on Microarchitecture, 2016.

[43] Sethumadhavan,S., Stolfo, S., Keromytis, A., Yang, J. "The SPARCHS Project", SysSec Workshop, The Netherlands, 2011.

[44] https://trustedcomputinggroup.org/trusted-platform-module-tpm-summary

[45] http://www.phoronix.com

[46] Intel, Intel 6 Series Chipset and Intel C200 Series Chipset, Data Sheet, May 2011

[47] AMD, AMD64 Architecture Programmer's Manual, Volume 2: System Programming.

[48] Lo, Tony C.S. "Tailoring TrustZone as SMM Equivalent", UEFI Plugfest, 2016. URL: http://www.uefi.org/sites/default/files/resources/UEFI_Plugfest_March_2016_AMI.pdf, accessed April 9, 2017.

[49] Intel, Intel Itanium Architecture Software Developer's Manual, Revision 2.3. Volume 2: System Architecture.

[50] https://github.com/f0rb1dd3n/papers/tree/master/rootkit_demonstration

[51] https://software.intel.com/en-us/blogs/2013/06/03/intel-xeon-phi-coprocessor-power-management-part-2a-core-c-states-the-details

[52] Gruss, D., Lipp, M. et al. "KASLR is Dead: Long Live KASLR", International Symposium on Engineering Secure Software and Systems, Bonn, Germany, 2017

# Appendix A: A complete example of EPA-RIMM's active monitoring phase

Figure 4 provides a worked example of Check decomposition, priority queue creation, bin formation, and measurement processing.

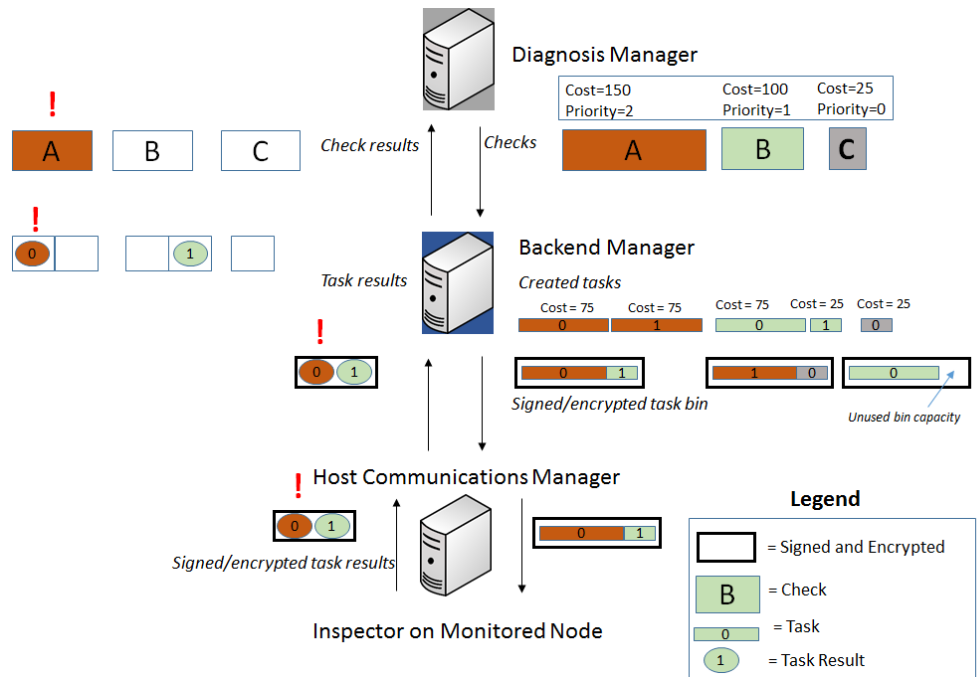

**Figure 4**. **A complete example of EPA-RIMM's active monitoring phase**. In this simplified example, the same Bin is provided to all monitored nodes, but in a heterogeneous environment the bins and the hash costs could differ between nodes. We show the Backend Manager and the Inspector as residing on separate machines, but there is no requirement for this separation. *(1) The DM creates three kernel function Checks: "A", "B", and "C" with costs of 150, 100, and 25 microseconds respectively, priorities of 2, 1, and 0 respectively (higher numbers indicate higher priority), and Decomposition targets of 75 microseconds; and sets an initial target Bin size of 100 microseconds. (2) The BEM performs decomposition and enqueues tasks A.0, A.1, B.0, B.1, and C.0. (3) The BEM forms the next Bin following the Backpack Algorithm, then signs and encrypts it. (4) The BEM send the Bin to the HCM. (5) The HCM triggers an SMI. (6) The HCM triggers an SMI. (7) The Inspector performs the measurements and sends the results to the Backend Manager. In this example, A.0 tests false, indicating an unexpected change. (8) The Backend Manager immediately passes this alert to the Diagnosis Manager. After the first Bin, Check B only has incomplete results and Check C has not been processed. When the bins containing the remaining tasks for Check B and C have been processed, the BEM merges the individual Task results into Check results which it provides to the Diagnosis Manager. In the event that the BEM does not receive a response from one of the bins within a predefined interval, the BEM will raise an alert to the Diagnosis Manager.*